\documentclass[aps,prl,showpacs,twocolumn,floats,epsfig,pdflatex]{revtex4}
\usepackage{amssymb}
\usepackage{amsbsy}
\usepackage{amsmath}
\usepackage{epsfig}
\usepackage{graphicx}
\begin{document}

\title {Superfluid-Insulator transition of ultracold atoms in an optical lattice
in the presence of a synthetic magnetic field}

\author{S. Sinha$^{(1)}$ and K. Sengupta$^{(2)}$}

\affiliation{$^{(1)}$Indian Institute of Science Education and
Research-Kolkata, Mohanpur, Nadia 741252, India.\\
$^{(2)}$Theoretical Physics Division, Indian Association for the
Cultivation of Sciences, Kolkata-700032, India. }

\date{\today}

\begin{abstract}

We study the Mott insulator-superfluid transition of ultracold
bosonic atoms in a two-dimensional square optical lattice in the
presence of a synthetic magnetic field with $p/q$ ($p$ and $q$ being
co-prime integers) flux quanta passing through each lattice
plaquette. We show that on approach to the transition from the Mott
side, the momentum distribution of the bosons exhibits $q$ precursor
peaks within the first magnetic Brillouin zone. We also provide an
effective theory for the transition and show that it involves $q$
interacting boson fields. We construct, from a mean-field analysis
of this effective theory, the superfluid ground states near the
transition and compute, for $q=2,3$, both the gapped and the gapless
collective modes of these states. We suggest experiments to test our
theory.

\end{abstract}


\maketitle

The physics of ultracold bosonic atoms in an optical lattice can be
well described by the Bose-Hubbard model \cite{bloch1, sachdev1}. In
fact, experiments on the Mott insulators-superfluid (MI-SF)
transitions of such bosonic atoms in two-dimensional (2D) optical
lattices \cite{spielman1} is found to agree with predictions of
theoretical studies of the Bose-Hubbard model quite accurately
\cite{fisher1,sengupta1,freericks1}. More recently, several
experiments have successfully generated time- or space- dependent
effective vector potentials for these neutral bosonic atoms by
creating temporally or spatially dependent optical coupling between
their internal states \cite{spielman2, spielman3}. Such a generation
of synthetic space-dependent vector potential and hence magnetic
fields is complementary to the conventional rotation technique
\cite{rotlat1}. Several theoretical studies have also been carried
on the properties of the bosons in an optical lattice in the
presence of an effective magnetic field \cite{ucpapers1}. In
particular, the MI-SF phase boundary has been computed using
mean-field theory \cite{lundhoker1} and excitation energy
calculation using a pertubative expansion in the hopping parameter
\cite{freericks2}. However, experimentally relevant issues such as
the momentum distribution of the bosons in the Mott phase, the
critical theory of the MI-SF transition, and the nature of the
superfluid ground states and collective modes near criticality have
not been addressed so far.

In this letter, we present a theory of the MI-SF transition for
ultracold bosons in a 2D square optical lattice with commensurate
filling $n_0$ and in the presence of a synthetic vector potential
corresponding to $p/q$ ($p$ and $q$ are co-prime integers) flux
quanta per plaquette of the lattice which addresses all of the
above-mentioned issues. The novel results of our work which have not
been addressed in earlier studies are as follows. First, using a
strong-coupling RPA theory for the bosons \cite{sengupta1}, we
provide an analytical formula for their momentum distribution in the
Mott phase and show that it develops $q$ precursor peaks on approach
to the MI-SF transition. Second, based on both the microscopic
strong-coupling theory and a symmetry analysis, we construct the
critical field theory for the transition and show that it
necessarily involves $q$ coupled boson fields \cite{balents1}.
Third, using a mean-field analysis of this effective theory, we find
the superfluid ground state to which the transition takes place and
chart out the corresponding spatial patterns of the superfluid
density. Finally, we compute the collective modes of the superfluid
phase for $q=2,3$, explicitly demonstrating the nature of both the
gapped and gapless collective modes near the transition, and provide
analytical expressions for their masses and group velocities in
terms of microscopic parameters of the theory. We suggest realistic
experiments which can verify specific predictions of our theory.

The Hamiltonian of a system of bosons in the presence of an optical
lattice and a synthetic magnetic field is given by \cite{bloch1,
fisher1,spielman1,lundhoker1,freericks2}
\begin{eqnarray}
{\mathcal H} &=& \sum_{{\bf r},{\bf r'}} J_{{\bf r}{\bf r'}} b_{{\bf
r}}^{\dagger} b_{{\bf r'}} + \sum_{{\bf r}} [-\mu {\hat n}_{{\bf r}}
+ \frac{U}{2} {\hat n}_{{\bf r}}({\hat n}_{{\bf r}}-1) ]
\label{ham1}
\end{eqnarray}
where $J_{{\bf r}{\bf r'}} = -J \exp(-i q^{\ast} \int^{\bf r'}_{\bf
r} \vec{A^{\ast}} \cdot \vec{dl}/ \hbar c)$, if ${\bf r}$ and ${\bf
r'}$ are nearest neighboring sites and zero otherwise,
$A^{\ast}=B^{\ast} (0,x)$ is the synthetic vector potential,
$q^{\ast}$($B^{\ast}$) is the effective charge (magnetic field) for
the bosons, $J$ is the hopping amplitude determined by the depth of
the optical lattice, and the value of $q^{\ast} B^{\ast}$ can be
controlled by varying the detuning between the hyperfine states of
the bosonic atoms \cite{spielman3}. Here $\mu$ is the chemical
potential, $U$ is the on-site Hubbard interaction, and $b_{\bf r}$
(${\hat n}_{\bf r}= b_{\bf r}^{\dagger} b_{\bf r}$) is the boson
annihilation (density) operator. In the rest of this work, we
consider the magnetic field to correspond to $p/q$ flux quanta
through the lattice: $q^{\ast}B^{\ast} a^2 /\hbar c = 2\pi p/q$, and
set the lattice spacing $a$, $\hbar$, and $c$ to unity.

The effect of the magnetic field manifests itself in the first term
of Eq.\ \ref{ham1} and thus vanishes in the local limit ($J=0$). In
this limit the boson Green function at $T=0$ can be exactly computed
\cite{sachdev1,sengupta1,freericks1}: $G_0(i\omega_n) = (n_0+1)(i
\omega_n - E_p)^{-1} -n_0(i \omega_n + E_h)^{-1}$. Here $\omega_n$
denote bosonic Matsubara frequencies and $E_h = \mu
-U(n_0-1)$($E_p=-\mu + Un_0$) are the energy cost of adding a hole
(particle) to the Mott state. To address the effects of the hopping
term, we write down the coherent state path integral corresponding
to ${\mathcal H}$: $Z= \int D \tilde \psi D \tilde \psi^{\ast}
\exp(-S)$ where $S = \int_0^{\beta} d\tau [(\sum_{{\bf r}} \tilde
\psi^{\ast}_{{\bf r}}(\tau)
\partial_{\tau} \tilde \psi_{\bf r}(\tau)  + {\mathcal H}[\tilde \psi^{\ast},
\tilde \psi])]$, $\tau$ is the imaginary time, $\beta=1/k_B T$ is
the inverse temperature ($T$), and $k_B$ is the Boltzman constant.
Following Ref.\ \cite{sengupta1}, we then decouple the hopping term
by two successive Hubbard-Stratonovitch transformations, integrate
out the original boson and the first Hubbard-Stratonovitch fields,
and obtain the final form of the strong-coupling effective action
$S_{\rm eff}= S_0 +S_1$
\begin{figure}
\rotatebox{0}{
\includegraphics*[width=0.8 \linewidth]{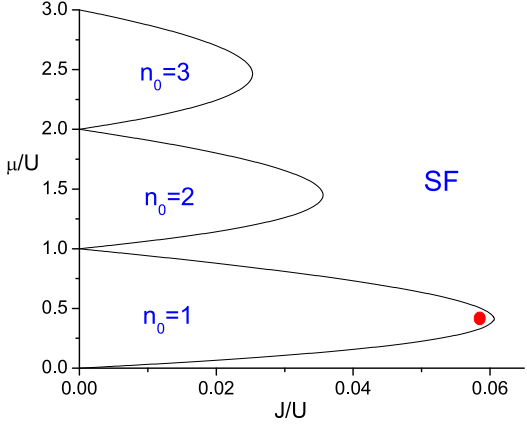}}
\caption{Color online) The MI-SF phase boundary for $q=2$. The red
dot indicates the value of $J$ and $\mu$ for which $n({\bf k})$ in
the left panel Fig. \ref{fig2} has been plotted.} \label{fig1}
\end{figure}
\begin{eqnarray}
S_0 &=& \int_{\bf k} \, \psi_{q}^{\ast} (i\omega_n, {\bf k})
[-G_0^{-1}(i\omega_n) I + J_q({\bf k}) ] \psi_{q} ( i \omega_n,{\bf
k}), \label{s0eq} \nonumber\\
S_1 &=& g/2 \int_0^{\beta} d \tau \int d^2 r |\psi_{q }^{\ast}({\bf
r},\tau) \psi_{q}({\bf r}, \tau)|^2, \label{s1eq}
\end{eqnarray}
where $\psi_q$ denotes the $q$-component auxiliary field introduced
through the second Hubbard-Stratonovich transformation and have the
same correlation functions as the original boson fields $\tilde
\psi$ \cite{sengupta1}, $\int _{\bf k} \equiv (1/\beta)
\sum_{\omega_n} \int d^2k/(2 \pi)^2$, $I$ denotes the unit matrix,
and $g>0$ is the static limit of the exact two-particle vertex
function of the bosons in the local limit \cite{sengupta1}. Here
$J_q({\bf k})$ is a $q\times q$ dimensional tridiagonal hermitian
matrix whose upper off-diagonal [diagonal] elements are $-J \exp(-i
k_y)$ [$-2J \cos(k_x + 2\pi \alpha/q)$], with $\alpha = 0,1,..q-1$.
It is well-known that $J_q({\bf k})$ has $q$ eigenvalues
$\epsilon_q^{\alpha}({\bf k})$ within the first magnetic Brillouin
zone ($-\pi\le k_y \le \pi$, $-\pi/q \le k_x \le \pi/q$) which are
$q$-fold degenerate. In particular, the lowest eigenvalue
$\epsilon_q^{m}({\bf k})$ has $q$ degenerate minima at ${\bf
Q}^{\alpha}=(0,2\pi\alpha/q)$ \cite{hof1}. Note that $S_0$
reproduces correct bosons propagator both in the local ($J=0$) and
the non-interacting ($U=0$) limits. Also, since $G_0^{-1}$ is
independent of momenta, finding the boson Green function
$G(i\omega_n,{\bf k})= [-G_0^{-1}(i\omega_n) I + J_q({\bf k})]^{-1}$
amounts to inverting $J_q({\bf k})$.

The critical hopping $J_c$ for the MI-SF transition as a function of
$\mu$ can be determined from the condition \cite{sachdev1}
\begin{eqnarray}
r_q &=& -G_0^{-1}(i\omega_n=0)+\epsilon_q^{m}({\bf k}={\bf
Q}^{\alpha})= 0.
\end{eqnarray}
The MI-SF phase boundary so obtained is shown in Fig.\ \ref{fig1}
for $q=2$ and agrees qualitatively with those obtained using
mean-field theory \cite{lundhoker1} and $J/U$ expansion
\cite{freericks2}. Note that $J_c$ remains same for all ${\bf
Q}^{\alpha}$ due to the $q$-fold degeneracy of $\epsilon_q^{m}({\bf
k})$.

\begin{figure}
\rotatebox{0}{
\includegraphics*[width=\linewidth,height=3.5cm]{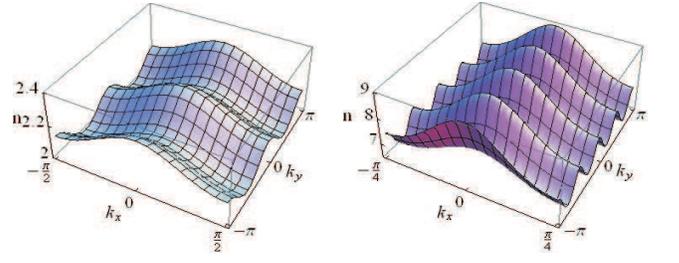}}
\caption{(Color online) Plot of $n({\bf k})$ for $q=2$ (left panel)
and $q=4$ ( right panel) at $\mu/U=0.414$ and $J/J_c=0.95$
indicating the precursor peaks in the Mott phase.} \label{fig2}
\end{figure}

The consequence of the $q$ fold degeneracy of $\epsilon^{\alpha}_q$
becomes evident in the momentum distribution of the bosons in the
Mott phase, which, at $T=0$, is given by $n({\bf k})= \lim_{T \to
0}(1/\beta) \sum_{\omega_n} {\rm Tr} G(i \omega_n, {\bf k})$. After
some straightforward algebra, one obtains
\begin{eqnarray}
n({\bf k})&=& \sum_{\alpha=0..q-1} \frac{E_q^{\alpha -} ({\bf k}) +
\delta \mu + U p}{E_q^{\alpha +}({\bf k})-E_q^{\alpha -}({\bf k})},
\label{momdist1}
\end{eqnarray}
where $\delta \mu= \mu-U(n_0-1/2)$, $p=(n_0+1/2)$ and $E_q^{\alpha
\pm}({\bf k}) =-\delta \mu + \epsilon_q^{\alpha}({\bf k})/2 \pm
\sqrt{\epsilon_q^{\alpha}({\bf k})^2 + 4 \epsilon_q^{\alpha}({\bf
k})U p + U^2}/2$ denotes the position of the poles of $G({\bf
k},i\omega_n)$ in the Mott phase. Note that $E_q^{\alpha}$ can also
be obtained from a time-dependent variational method \cite{sinha1}.

Eq.\ \ref{momdist1} is a central result of this work and generalizes
its counterpart in Ref.\ \onlinecite{sengupta1} in the presence of a
magnetic field. The peaks of $n({\bf k})$ occur when $E_q^{\alpha
+}({\bf k})-E_q^{\alpha -}({\bf k})$ becomes small near the MI-SF
transition. The degeneracy of $\epsilon_q^{\alpha}({\bf k})$ and
hence $E_q^{\alpha \pm}({\bf k})$ ensures that this happens at $q$
points in the first Brillouin zone leading to $q$ precursor peaks in
$n({\bf k})$ at ${\bf k}={\bf Q}^{\alpha}$. This is demonstrated in
Fig. \ref{fig2} for $q=2$ and $q=4$. Note that the positions of
these peaks in the Brillouin zone depend on the specific form of the
vector potential realized in the experiments; for symmetric vector
potentials generated by rotation they would appear at $(\pi
\alpha/q,\pi \alpha/q)$. However, their number depends only on $p/q$
and the lattice geometry.

At $J_c$, the MI-SF transition occurs since the energy gap to
addition of particles and/or holes to the Mott state vanishes. In
contrast to standard superfluid-insulator transition
\cite{fisher1,sengupta1,freericks1}, the presence of $q$ degenerate
minima at ${\bf k}={\bf Q}^{\alpha}$ necessitates the corresponding
Landau-Ginzburg theory to be constructed out of $q$ low-energy
fluctuating fields $\phi^{\alpha} ({\bf r},t)$ around these minima:
\begin{eqnarray}
\psi_q ({\bf r},t) &=& \sum_{\alpha=0 .. q-1} \chi^{\alpha}_q ({\bf
r}) \phi^{\alpha}({\bf r},t), \label{field1}
\end{eqnarray}
where $\chi^{\alpha}_q({\bf r})$ denotes the eigenvectors of
$J_q({\bf Q}^{\alpha})$ in real space, and we have Wick-rotated to
real time. The quadratic part of the Landau-Ginzburg theory,
obtained by expanding $S_0$ (Eq.\ \ref{s0eq}) about the minima, is
given by
\begin{eqnarray}
{\mathcal S}_0 &=& \int d^2 r dt \sum_{\alpha=0..q-1} \phi^{\alpha
\ast}({\bf r},\tau) \big[ K_0
\partial_{t}^2 + i K_1
\partial_{t} \nonumber\\
&& +r_q - v_q^2 (\partial_x ^2 + \partial_y^2) \big]
\phi^{\alpha}({\bf r},\tau), \label{quad1}
\end{eqnarray}
where $K_0 = 1/2 \partial^2 G_0^{-1}/\partial \omega^2|_{\omega=0} =
n_0(n_0+1)U^2/(\mu+U)^3$, $K_1=
\partial G_0^{-1}/\partial \omega|_{\omega=0} = 1 -
n_0(n_0+1)U^2/(\mu+U)^2$, and $v_q^2 = \nabla^2_{{\bf k}}
\epsilon^m_q({\bf k})/2$ with $v_2^2 = J/\sqrt{2}$. At the tip of
the Mott lobe, where $\mu= \mu_{\rm tip} =U(\sqrt{n_0(n_0+1)}-1)$,
$K_1=0$. Thus we have a critical theory with dynamical critical
exponent $z=1$. Away from the tip, $K_1 \ne 0$ rendering $z=2$
\cite{sachdev1}.

\begin{figure}
\rotatebox{0}{
\includegraphics*[width=\linewidth]{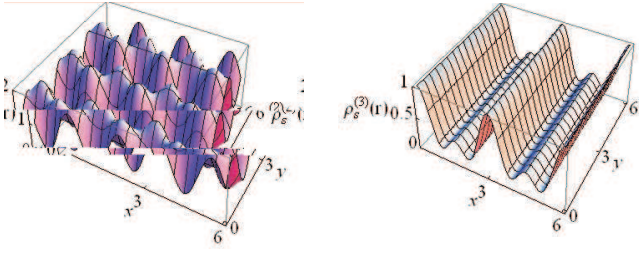}}
\caption{(Color online) Plot of the normalized superfluid density
$\rho_s^{(2)}$ for $q=2$ (left panel) and $\rho_s^{(3)}$ for $q=3$
(right panel). } \label{fig3}
\end{figure}

The most general quartic Landau-Ginzburg action in terms of $q$
bosonic fields which is allowed by invariance under projective
symmetry group (PSG) of the underlying square lattice has been
obtained in Ref.\ \cite{balents1}. The elements of the PSG for the
square lattice include translation along $x$ and $y$, rotation by
$\pi/2$ about $z$ axis, and reflections about $x$ and $y$ axes. The
transformation properties of $\phi^{\alpha}$ fields under these
operations are tabulated in Ref.\ \cite{balents1}. The invariant
quartic action so obtained is given by ${\mathcal S}_1 = \int d^2 r
dt \sum_{\alpha,\beta,\gamma=0}^{q-1} \Gamma_q^{\beta \gamma}
\phi^{\alpha \ast} \phi^{\alpha+\beta \ast} \phi^{\alpha+\gamma}
\phi^{\alpha+\beta-\gamma}/4$, where $\Gamma_q^{\alpha
\beta}=\Gamma_q^{-\alpha -\beta}= \Gamma_q^{\alpha-\beta
\,\beta}=\Gamma_q^{\alpha-2\beta \, -\beta}$ and sums over integers
$\alpha$, $\beta$, and $\gamma$ are taken modulo $q$. Eq.
\ref{quad1} along with ${\mathcal S}_1$ has been analyzed in details
in Ref.\ \cite{balents1}. However, the lack of microscopic knowledge
of $\Gamma_q^{\alpha \beta}$ did not allow identification of the
exact ground state of ${\mathcal S}_1$; only possible
symmetry-allowed ground states were charted.

\begin{figure}
\rotatebox{0}{
\includegraphics*[width=0.8\linewidth]{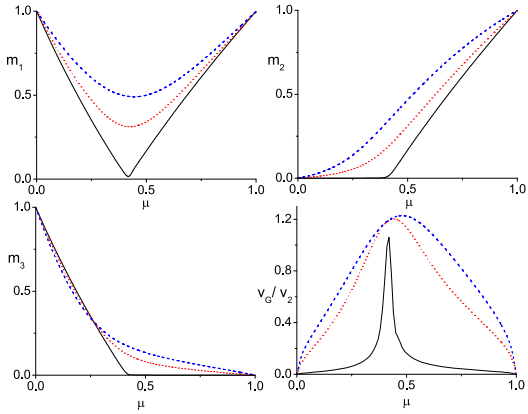}}
\caption{(Color online) Top left, right, and bottom left panels:
Plot of $m_{1,2,3}$ as a function of $\mu$ for $J/J_c =1$ (solid
black line), $1.2$ (red dotted line) and $1.5$ (blue dashed line).
Bottom right panel: Plot of $v_G/v_2$ vs $\mu$ for $J/J_c =1.01$
(black solid line), $1.2$ (red dotted line) and $1.5$ (blue dashed
line). $U=1$, $n_0=1$, $q=2$, and $\mu_{\rm tip}= 0.414$ for all the
plots. } \label{fig4}
\end{figure}

Here, taking advantage of the microscopic knowledge of $g$ and
$\chi^{\alpha}_{\bf r}$, we determine the exact superfluid state to
which the transition takes place. This is done by substituting of
Eq.\ \ref{field1} in Eq.\ \ref{s1eq} followed by coarse-graining of
the resultant action which involves replacing $\chi^{\alpha}_{\bf
r}$ by its sum over $q$ lattice sites: $\int d^2 r dt
L_1[\chi_q^{\alpha}({\bf r})] L_2[\phi^{\alpha}({\bf r},t)] \to
\{(1/q^2) \sum_{x,y=0}^{q-1} L_1[\chi_q^{\alpha}({\bf r})] \} \int
d^2 r dt L_2[\phi^{\alpha}({\bf r},t)] = c_0 \int d^2 r $ $dt\,
L_2[\phi^{\alpha}({\bf r},t)]$. Here $L_1$ and $L_2$ denotes
arbitrary fourth order polynomial functions and the coarse-graining
procedure is applicable due to the natural separation of scale
between the spatial variation of $\chi^{\alpha}_q({\bf r})$ and
$\phi^{\alpha}({\bf r},t)$. The effective action so obtained is then
compared to ${\mathcal S}_1$ to obtain $\Gamma_q^{\alpha \beta}$.
Finally, we minimize the resultant action at the mean-field level
and obtain the superfluid ground state near the MI-SF transition.
This procedure is most easily demonstrated for $q=2$. Here,
$\epsilon_2^m ({\bf k}) = -2 J \sqrt{ \cos^2(k_x) + \cos^2(k_y)}$
leading to two minima at $(k_x,k_y)=(0,0)\,{\rm and}\,(0,\pi)$ with
eigenfunctions $\chi_2^{0}({\bf r}) = (1+ \sqrt{2} + \exp(i\pi
x))/\sqrt{4+2\sqrt{2}}$ and $\chi_2^{1}({\bf r}) = \exp(i \pi y)(1+
\sqrt{2} - \exp(i\pi x))/\sqrt{4+2\sqrt{2}}$. Putting these values
in Eq.\ \ref{field1} and Eq.\ \ref{s1eq}, the coarse-grained
effective action reads $S_{\rm eff}^{q=2} = 1/8 \int d^2r dt
[3g(|\phi^0({\bf r},t)|^2 +|\phi^1({\bf r},t)|^2)^2 + g (\phi^{0
\ast}({\bf r},t)\phi^{1}({\bf r},t)-\phi^{1\ast}({\bf
r},t)\phi^{0}({\bf r},t))^2 ]$. Comparing $S_{\rm eff}^{q=2}$ with
${\mathcal S}_1$ for $q=2$, we find $\Gamma_2^{00}= 3g/2$ and
$\Gamma_2^{10}=g/2$. A mean-field analysis then yields the
superfluid ground state: $\langle \phi^0({\bf r},t)\rangle = \phi^0
= i \phi^1 = \langle \phi^1({\bf r},t)\rangle$. The renormalized
superfluid density can be obtained by using $ \rho_s^{(2)}({\bf r})
= |\psi_2^{\rm MF}({\bf r})|^2/|\psi_2^{\rm MF}(0)|^2$ where
$\psi_2^{\rm MF}$ is obtained by substituting $\langle
\phi^{0,1}({\bf r},t)\rangle $ in Eq.\ \ref{field1}. Analogous
procedure carried out for $q=3$ yields the superfluid ground state:
$\langle \phi^{0}({\bf r},t) \rangle \ne 0, \, \langle \phi^{\alpha
\ne 0}({\bf r},t) \rangle =0$. The resultant plots of
$\rho_s^{(q)}({\bf r})$, shown in Fig.\ \ref{fig4} for $q=2$, and
$q=3$, display  $2$ and $3$ sublattice patterns respectively. We
note that the procedure mentioned above constitutes a general method
for obtaining the superfluid ground state and density near the MI-SF
critical point for any $q$.

Finally, we compute the collective modes of the superfluid ground
state near the transition. First we consider the case $q=2$ and
rewrite $S_{\rm eff}^{(2)}$ in terms of a linear combination of the
$\phi^{\alpha}$ fields: $\xi^{0[1]} = (\phi^{0}+[-]i
\phi^{1})/\sqrt{2}$. The quartic action becomes $S_{\rm
eff}^{'\,q=2} = 1/8 \int d^2r dt [3g (|\xi^0({\bf r},t)|^2
+|\xi^1({\bf r},t)|^2)^2 - g (|\xi^{0}({\bf r},t)|^2-|\xi^{1}({\bf
r},t)|^2)^2]$ so that the superfluid ground state corresponds to
condensation of $\xi^{0}$. The quadratic action can be written as
${\mathcal S}_0^{'} = \int d^2 r d t \sum_{\alpha=0}^{q-1}
\xi^{\alpha \ast}({\bf r},t) [-G_0^{-1}(\omega) - c_2 + v_2^2 |{\bf
k}|^2] \xi^{\alpha}({\bf r},t)$, with $c_2=-\epsilon_2({\bf
k}=0)=2\sqrt{2} J $. Using these actions, and carrying out a
straightforward linearization $\xi^0({\bf r},t)=\xi^{0} + \delta
\xi^{0}({\bf r},t)$ and $\xi^1({\bf r},t)= \delta \xi^{1}({\bf
r},t)$, where $\xi^0= \sqrt{2|r_2|/g}$, we find that there are four
collective modes. Two of these correspond to the condensed field
$\xi^0$ and $\xi^{0 \ast}$, and have dispersions
\begin{eqnarray}
\omega^{(1)}_{\pm} =(\pm B_2({\bf k})/2 + \left[B_2^2({\bf k})/4-
C_2({\bf k}) \right]^{1/2})^{1/2}, \label{coll1}
\end{eqnarray}
where  $B_2({\bf k})= [2 \delta \mu - A_2({\bf k})]^2 + 2 \alpha_0
(v_2^2 |{\bf k}|^2 -r_2)-r_2^2$, $C_2({\bf k})= \alpha_0^2 v_2^2
|{\bf k}|^2 ( v_2^2 |{\bf k}|^2  -2 r_2)$, $\alpha_0=(U+\mu)$, and
$A_2({\bf k})= -c_2 + v_2^2 |{\bf k}|^2 -2 r_2$. At low wave-vector,
$\omega^{(1)}_+$ is gapped with a mass $m_1= \sqrt{B_2(0)}$ while
$\omega^{(1)}_-$ has linear dispersion with velocity $v_G= v_2
\alpha_0 \sqrt{2|r_2|}/m_1$. The other two modes, which correspond
to the non-condensed field $\chi^1$ and $\chi^{1 \ast}$, have
dispersions
\begin{eqnarray}
\omega^{(2)}_{\pm} = \pm D_2({\bf k})/2 + [D_2({\bf k})^2/4 +
\alpha_0 (|r_2|/2 + v_2^2 |{\bf k}|^2)]^{1/2}, \nonumber
\end{eqnarray}
where $D_2({\bf k})= -(2 \delta \mu + c_2 - v_2^2 |{\bf k}|^2
+r_2/2)$. Both these modes are gapped in the superfluid phase with
masses $m_{2[3]}= +[-]D_2(0)/2 +\sqrt{D_2(0)^2/4 +
\alpha_0|r_2|/2}$. The masses $m_{1,2,3}$ and the velocity $v_G$ of
these modes, plotted as a function $\mu$ in Fig.\ \ref{fig4} for
several representative values of $J/J_c$, displays the following
characteristics. At $\mu=\mu_{\rm tip}$ and $J=J_c$, where $2 \delta
\mu= - c_2$ rendering $B_2(0)=0$ and $D_2(0)=0$, all the modes
become gapless with $\omega \sim |{\bf k}|$ dispersion. Also at $\mu
\ne \mu_{\rm tip}$, one of the two modes $\omega^{(2)}_{\pm}$ always
remain gapless at $J=J_c$ with $\omega \sim |{\bf k}|^2$ dispersion.
The velocity $v_G$ at $J=J_c$, is non-zero only at $\mu=\mu_{\rm
tip}$; thus it shows a peak at $\mu_{\rm tip}$ for $J$ close to
$J_c$. We emphasize that our theory specifies $v_G$ and $m_{1,2,3}$
in terms of the parameters of the Bose-Hubbard model.

For $q=3$ only $\phi^0$ condense, and the corresponding collective
modes are given by Eq.\ \ref{coll1} with $c_2 , v_2, r_2 \to c_3,
v_3, r_3$ (where $c_3=-\epsilon^m_3(0)$). This leads to similar
gapped and a gapless mode with linear dispersion as for $q=2$.
However, the dispersion of the non-condensed modes are different.
The effective action $S_{\rm eff}^{'\,q=3}$ turns out to be $O(3)$
symmetric: $S_{\rm eff}^{'\,q=3} \sim \int d^2r dt
(\sum_{\alpha=0..2}| \xi^{\alpha}({\bf r},t)|^2)^2$ leading to two
doubly-degenerate non-condensed modes $\omega^{(3)}_{\pm} =(\pm
D_3({\bf k}) + \sqrt{D_3({\bf k})^2 + v_3 |{\bf k}|^2})/2$, where
$D_3({\bf k})= -(2 \delta \mu + c_3 - v_3^2 |{\bf k}|^2)$. Thus
there are two gapped and two gapless modes with $\omega \sim |{\bf
k}|^2$. These two modes become gapless due to the $O(3)$ symmetric
form of $S_{\rm eff}^{'\,q=3}$. For $q>3$, there are in general $2q$
collective modes, and we have left their analysis as a subject of
future study.

For experimental verification of our theory, we suggest measurement
of $n({\bf k})$ for the bosons in the Mott phase near the transition
as done earlier in Ref.\ \cite{spielman1} for 2D optical lattices
without the synthetic magnetic field. This distribution is predicted
to display $q$ precursor peaks. The collective modes in the
superfluid phase can also be directly probed and compared to the
theory by standard lattice modulation experiments \cite{esslinger1}
and response functions measurement by Bragg spectroscopy
\cite{ketterle1}.

In conclusion, we have analyzed the MI-SF transition of ultracold
bosons in a 2D optical lattice in the presence of a synthetic
magnetic field. We have demonstrated the presence of $q$ precursor
peaks in their momentum distribution near the MI-SF transition,
provided a critical field theory for the transition, analyzed this
theory to predict the ground state and the collective modes of the
bosons in the superfluid phase, and suggested experiments to probe
our theory. K.S. thanks DST, India for financial support under
Project No. SR/S2/CMP-001/2009.


\vspace{-0.7 cm}





\begin{thebibliography}{99}

\bibitem{bloch1} M. Greiner, {\it et al.}, Nature {\bf 415}, 39 (2002);
C. Orzel {\it et al.}, Science {\bf 291}, 2386 (2001).

\bibitem{sachdev1} S. Sachdev, {\it Quantum Phase Transitions},
(Cambridge University Press, Cambridge, England, 1999), Chap. 11.

\bibitem{spielman1} I. B. Spielman, W. D. Phillips, and J. V. Porto,
\prl {\bf 98}, 080404 (2007).

\bibitem{fisher1} M. P. A. Fisher {\it et al.}, Phys. Rev. B {\bf 40}, 546 (1989);
K. Seshadri {\it et al.}, Europhys. Lett. {\bf 22}, 257 (1993); D.
Jaksch {\it et al.}, Phys. Rev. Lett. {\bf 81}, 3108 (1998).

\bibitem{sengupta1} K. Sengupta and N. Dupuis, \pra {\bf 71}, 033629
(2005).

\bibitem{freericks1} J. Freericks {\it et al.}, \pra {\bf 79}, 053631
(2009).

\bibitem {gaugepapers1}D. Jaksch and P. Zoller, New J. Phys. {\bf 5},
56 (2003); E. Mueller, Phys. Rev. A {\bf 70}, 041603(R) (2004); K.
Osterloh {\it et al}, Phys. Rev. Lett. {\bf 95}, 010403 (2005); N.
Goldman {\it et al.} \pra {\bf 79}, 023624 (2009); I. B. Spielman,
Phys. Rev. A {\bf 79}, 063613 (2009).

\bibitem{spielman2} Y-J. Lin {\it et al}, Phys. Rev. Lett. {\bf 102},
130401 (2009).

\bibitem{spielman3}Y.-J. Lin {\it et al}, Nature {\bf 462}, 628-632
(2009).

\bibitem{rotlat1} I. Coddington {\it et al}, \prl {\bf 91}, 100402
(2003); A. Aftalion, X. Blanc, and J. Dalibard, Phys. Rev. A {\bf
71}, 023611 (2005).

\bibitem{ucpapers1} N. Goldman {\it et al.}, Phys. Rev. Lett. {\bf 103},
035301 (2009); I. Satija, D. C. Dakin and C. W. Clark, Phys Rev
lett, {\bf 97}, 216401, (2006); S-L Zhu {\it et al.}, Phys. Rev.
Lett. {\bf 97}, 240401 (2006); X-J Liu {\it et al.}, Phys. Rev.
Lett. {\bf 98}, 026602 (2007).

\bibitem{lundhoker1} R. O. Umucalilar and M. O. Oktel, Phys. Rev. A {\bf 76},
055601 (2007); E. Lundh, EuroPhys. Lett. {\bf 84}, 10007 (2008).

\bibitem{freericks2} M. Niemeyer, J. K. Freericks, H. Monien, Phys. Rev. B {\bf 60},
2357 (1999).

\bibitem{balents1} L. Balents {\it et al.}. Physical Review B {\bf 71},
144508 (2005).

\bibitem{hof1} D. Hofstadter, Phys. Rev. B {\bf 14}, 2239
(1976); M. Kohmoto, \prb {\bf 39}, 11943 (1989).

\bibitem{sinha1} D. L. Kovrizhin, G. V. Pai, and S. Sinha,
arXiv:0707.2937 (unpublised).

\bibitem{esslinger1} R. Jordens {\it et al.}, Nature {\bf 455}, 204 (2008).

\bibitem{ketterle1} J. Stenger {\it et al.}, Phys. Rev. Lett. {\bf 82},
4569 (1999).

\end{thebibliography}
\end{document}